# Multifunctional Charge Transfer Plasmon Sensors

Alemayehu Nana Koya* and Wei Li*

*GPL Photonics Laboratory, State Key Laboratory of Luminescence and Applications, Changchun Institute of Optics, Fine Mechanics and Physics, Chinese Academy of Sciences, Changchun 130033, Jilin, P. R. China*

E-mail: alemayehukoya@ciomp.ac.cn; weili1@ciomp.ac.cn

**Abstract**

Charge transfer plasmon (CTP) modes arise when metallic nanoparticles are connected by a conductive junction. These ultatunable plasmonic modes can be actively tuned and broadly modulated from visible to IR regimes, implying their potentials for applications in sensing. This review showcases recent developments in theory and applications of charge transfer plasmon resonances (CTPRs) in various configurations of conductively linked plasmonic nanodimers. In particular, we give a due attention to exploiting ultratunable spectral features of charge transfer plasmon resonances for developing multifunctional plasmonic sensors for bulk, surface, gas and molecular sensing applications. We have also briefly discussed that the charge and energy transfer between two plasmonic nanoparticles linked by sub-nanometer length self-assembled monolayers are of interest for single-molecule detection and molecular electronics. In addition to the well-established surface plasmon resonance and localized surface plasmon resonance based sensing schemes, CTPR sensors may open up a new route in efforts to develop multifunctional sensing technologies.

## 1. Introduction

The interaction of light with nanostructured noble metals gives rise excitation of surface plasmon resonance – a collective oscillation of free electrons in the conduction band of metallic nanostructures. [1] Plasmon resonance is associated with exotic photo-induced phenomena such as strong resonances, hugely enhanced and extremely confined near fields, and phtothermal heating, [2, 3] which have been widely exploited for different applications including enhanced spectroscopy, [4] photocatalysis, [5] plasmon-enhanced optical trapping, [6] single-molecule studies, [7] and sensing. [8] The coherent excitation of surface plasmons can take various forms including surface plasmon resonance (SPR) at metal-dielectric interface, localized surface plasmon resonance (LSPR) in metallic nanoparticles, or charge transfer plasmon resonance in connected metallic nanostructures. [9] These resonance properties metallic nanostructures are highly dependent on the size, shape, configuration, and material composition of the nanostructures and the refractive index of the surrounding medium. [10]

In particular, the charge transfer plasmon modes that arise in conductively bridged plasmonic nanoparticles can be broadly tuned from visible to near-infrared regime by manipulating the nanojunction conductance. [11, 12] The emergence of these low-energy modes has been observed in several configurations including metallic nanodimers, [13] nanoshells, [14] and particle nanochains. [15] One can also observe CTP-like modes in nearly touching plasmonic nanodimers having atomic scale inter-particle separations. [16, 17] With actively tunable spectral features and versatile geometries, thus, CP modes in plasmonic nanosystems are of interest for devising ultrafast optoelectronic devices and ultrasensitive sensors. As a result, charge transfer plasmon modes have been extensively explored by a number of researchers with the aim of understanding the underlying physics of such modes and exploiting their potentials for various applications. [18, 19]

On the other hand, the plasmonic sensing schemes developed so far have been solely based on the SPR and LSPR. [20 - 22] In conductively linked plasmonic nanosystems, one can observe multiple modes including bonding dimer plasmon (BDP) type mode, ultratunable CTP mode, and dipole-CTP hybrid mode. [23] Given interesting spectral signatures of the charge transfer plasmon resonances, high quality factor of the dipole-CTP mode, and versatile geometric parameters of linked plasmonic nanostructures, the sensing technology can benefit a lot from charge transfer plasmon resonance. In fact, the CTPR-based sensing principle is as same as that of the LSPR-based sensing except that the former has versatile geometries and broadly tunable spectral features. As schematically illustrated in **Figure 1**, CTPR sensors can be employed for sensing various samples ranging from bulk materials to single-molecules and gas molecules.

Within this framework, here we overview the recent advances in theory and applications of charge transfer plasmon resonances in conductively linked plasmonic nanosystems. In particular, we showcase current developments in theory of charge transfer plasmons in conductively linked plasmonic nanoparticles with interparticle separations ranging from nanometer scale to atomistic regime. The charge transfer across the conductive junction can be actively controlled using various materials including metals,[24] palladium (Pd) nanoparticles,[25] hyphotetical molecules,[26] DNA linkers,[27] and self-assembled monolayers (SAMs). [28] Moreover, we give a particular attention to exploitation of CTP resonances for sensing applications that include bulk sensing, surface sensing, molecular sensing, and gas sensing. We also showcase quantum plasmon resonances for single-molecule detection using single-molecule break junction techniques. Thus, we anticipate that these ultratunable CTPR-based sensors may lead to a paradigim shift in plasmonic sensing.

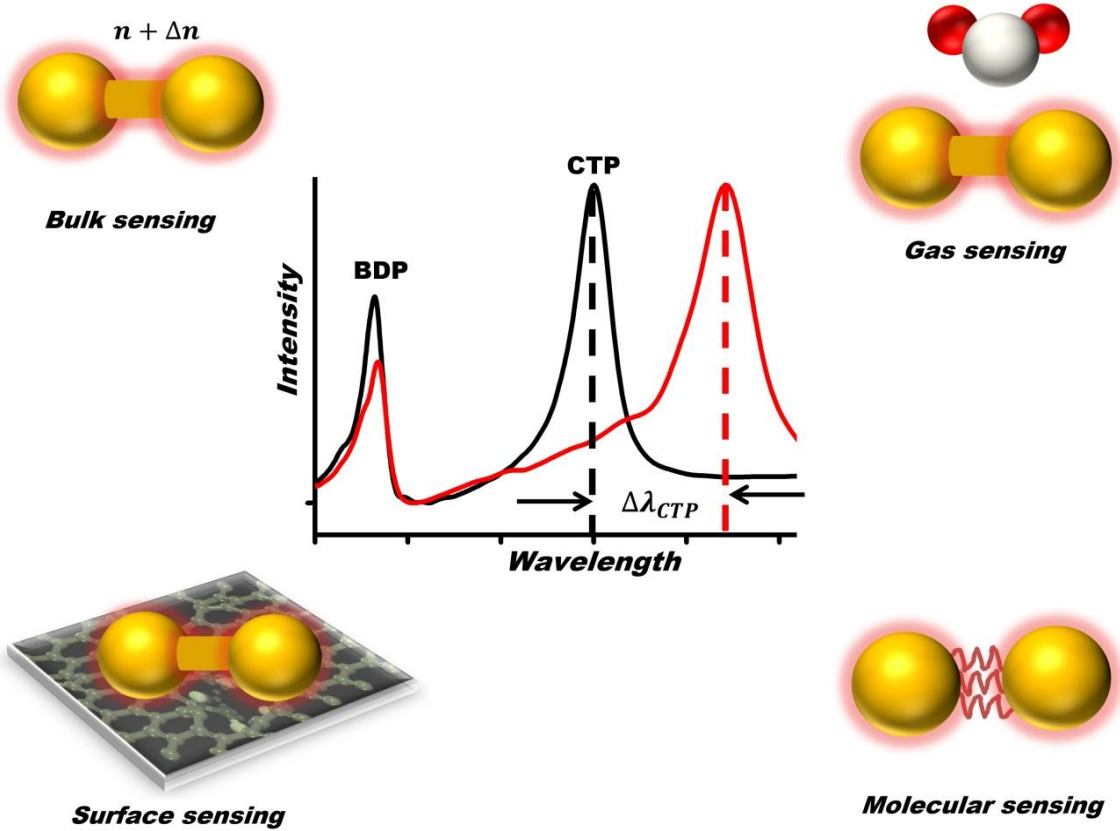

**Figure 1**. Illustration of sensing principles based charge transfer plasmon resonance for refractive index sensing (top left), gas sensing (top right), surface sensing (bottom left), and molecular sensing (bottom right). Middle: CTP resonance wavelength shift ($\Delta\lambda_{CTP}$) due to change in the dielectric environment of linked nanoparticle dimer.

## 2. Classical theory of charge transfer plasmon resonances

The electromagnetic coupling of plasmonic resonances in interacting metallic nanoparticles offers the opportunity to excite exotic plasmonic modes like charge transfer plasmon resonances in conductively linked metallic nanoparticles. For a nanoparticle dimer connected by conductive junction with nanometer scale length and thickness, one can describe the charge flow through the junction using classical electrodynamics. Generally, the conductance $G$ of cylindrical nanojunction having radius $r$ and length $L$ (see **Figure 2a**) can be expressed as.[12]

$$G = \frac{\sigma \pi r^2}{L} \tag{1}$$

where $\sigma$ is the conductivity of the nanojunction. Nevertheless, the onset of CTP modes in conductively linked plasmonic nanoparticles depends not only on the junction conductance as shown in **Equation 1** but also on the size of the linked nanoparticles. One can determine the conductance threshold for the onset of CTP resonance in linked plasmonic nanosystems with [29, 30]

$$G_{CTP} = \frac{c}{8\lambda_{CTP}} \frac{D^2}{L} \qquad (2)$$

where $\lambda_{CTP}(= 2\pi c/\omega_{CTP})$ is the resonance wavelength of charge transfer plasmon and *D* is the diameter of spherical nanoparticle as illustrated in **Figure 2a.** When the minimum requirement implied in **Equation 2** is met, one can observe CTP resonance, which appears at longer wavelength compared to the bonding dimer plasmon (BDP) mode (**Figure 2b**).

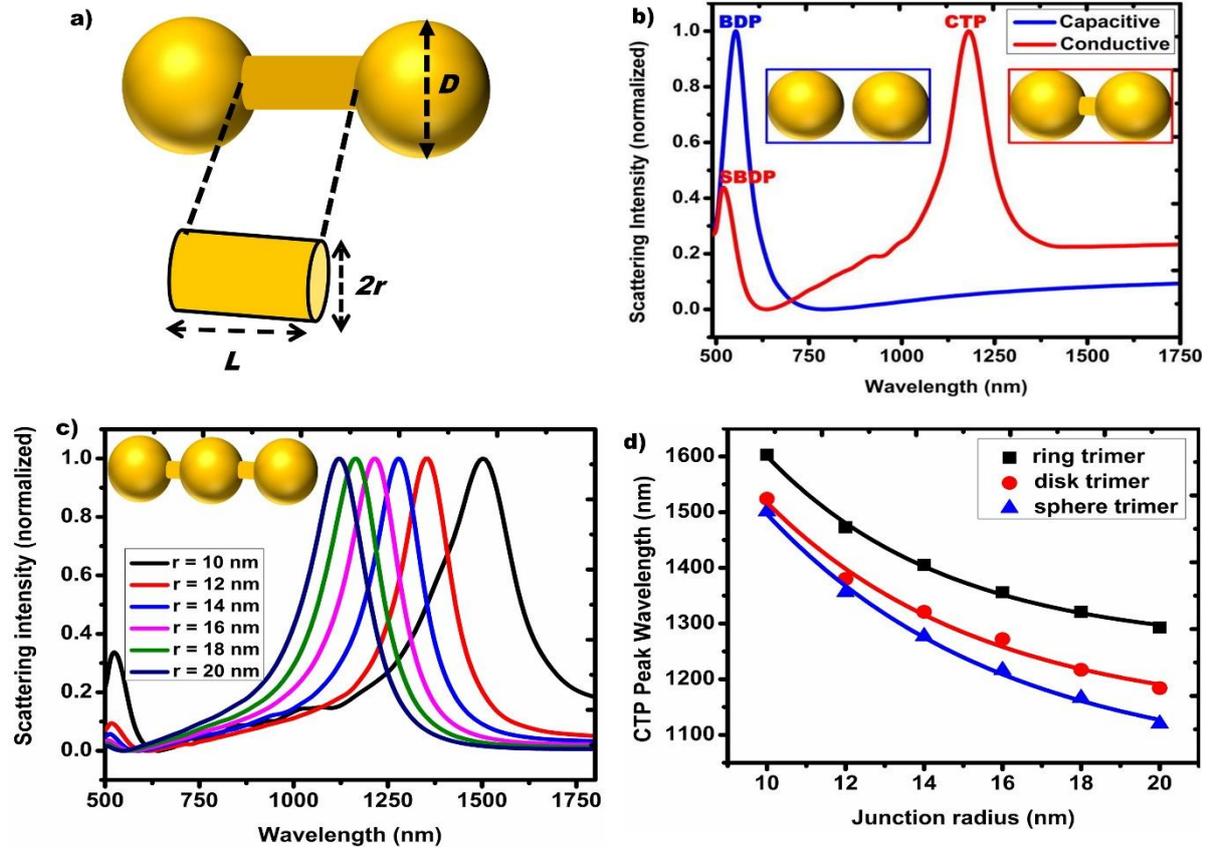

**Figure 2**. Spectral characteristics of charge transfer plasmon resonances in conductively linked plasmonic (here Au) nanoparticles. (a) Schematic illustration of geometric parameters that define CTP resonances in conductively linked plasmonic nanoparticles. These include nanoparticle size *D*, junction length *L* and radius *r*. (b) Typical spectral feature of the charge transfer plasmon resonance of connected Au nanoparticles. The CTP mode appears substantially at lower energies compared to other plasmon modes like screened bonding dimer plasmon (SBDP) mode of bridged nanodimer or bonding dimer plasmon (BDP) mode of unbridged nanodimer. Reproduced with permission from ref.[18] Copyright 2017 AIP Publishing. (c) Tuning CTP resonance of linked Au nanotrimer by changing junction conductance through its geometric parameter (here radius of cylindrical linker). (d) CTP spectral shift of conductively linked Au trimers as functions of junction radius and nanoparticle shapes. Reproduced with permission from ref .[12] Copyright 2016 AIP Publishing.

Extending this fact to more complex geometries and various shapes, Koya and Lin developed a generalized analytical expression for charge transfer plasmon resonance shift as functions of junction radius *r* and nanoparticle shape.[12, 18]

$$\lambda_{CTP} = \lambda_{CTP}^0 + \frac{C}{exp\left[\frac{r-r_0}{t}\right]} \tag{3}$$

where $\boldsymbol{\lambda_{CTP}^0}$ is the CTP resonance wavelength at the smallest junction radius (*r* = 10 nm in this case), and *C*, *r₀* and *t* are shape-specific fitting parameters. As implied in **Equation 3**, increasing the junction radius leads to a blueshift of the CTP wavelength (**Figure 2c**). As shown in **Figure 2d**, the CTP spectral shift depends not only on the nanojunction conductance but also on the size and shape of the nanoparticles linked by the junction. This is attributed to the excitation of sub-radiant and super-radiant plasmon modes that arise in complex nanostructures like nanoring and nanodisk dimers. [31, 32] Generally, the CTP resonance shift of coupled plasmonic nanodimers can be modeled with the classical electrodynamics.[33, 34] However, when the size of interparticle separation enters to sub-nanometer scale, the universal scaling relationship fails to describe the spectral shift, as quantum mechanical effects emerge in such atomistic scale gaps. In the next section, we cover these and other issues of conductively linked plasmonic nanodimers with sub-nanometer scale inter-particle separations.

## 3. DNA-assisted charge transfer in plasmonic nanodimers

When plasmonic nanoparticles are linked by narrow conductive molecular junctions with sub-nm scale dimension, the classical electrodynamics fails to accurately describe the plasmonic coupling, as non-local screening and electron tunneling effects become significant. [35] To fill this gap, Esteban *et al*. developed quantum-corrected model (QCM), which incorporates quantum-mechanical effects within a classical electrodynamic framework. [36] QCM has been widely employed to describe charge transfer in molecularly linked plasmonic nanodimers with sub-nm gaps, where one can actively control the charge transfer across the conductive junction by modeling its conductance with Pd and DNA molecules (see **Figure 3a**). [37, 38]

In this regard, Lerch and Reinhard explored the role of DNA mediated charge transfer in plasmonic nanodimers, experimentally demonstrating DNA mediated optical tunneling in Au nanodimers with separations as long as 2.8 nm.[27] In related work, they investigated the effect of interstitial nanoparticles on the distance-dependent plasmon coupling between DNA-connected Au nanoparticles.[25] In particular, they studied the effect of increasing Pd nanoparticles intragap density on the optical response of DNA-tethered plasmonic nanodimers (see **Figure 3b**). They demonstrated that high Pd densities in the inter-particle gap increases the gap conductance and induces the transition from capacitive coupling to conductive coupling (**Figure 3c&d**). Pd was chosen since its cations are best known to bind to DNA and Pd–Au heterostructures have interesting implications for sensing applications. [25]

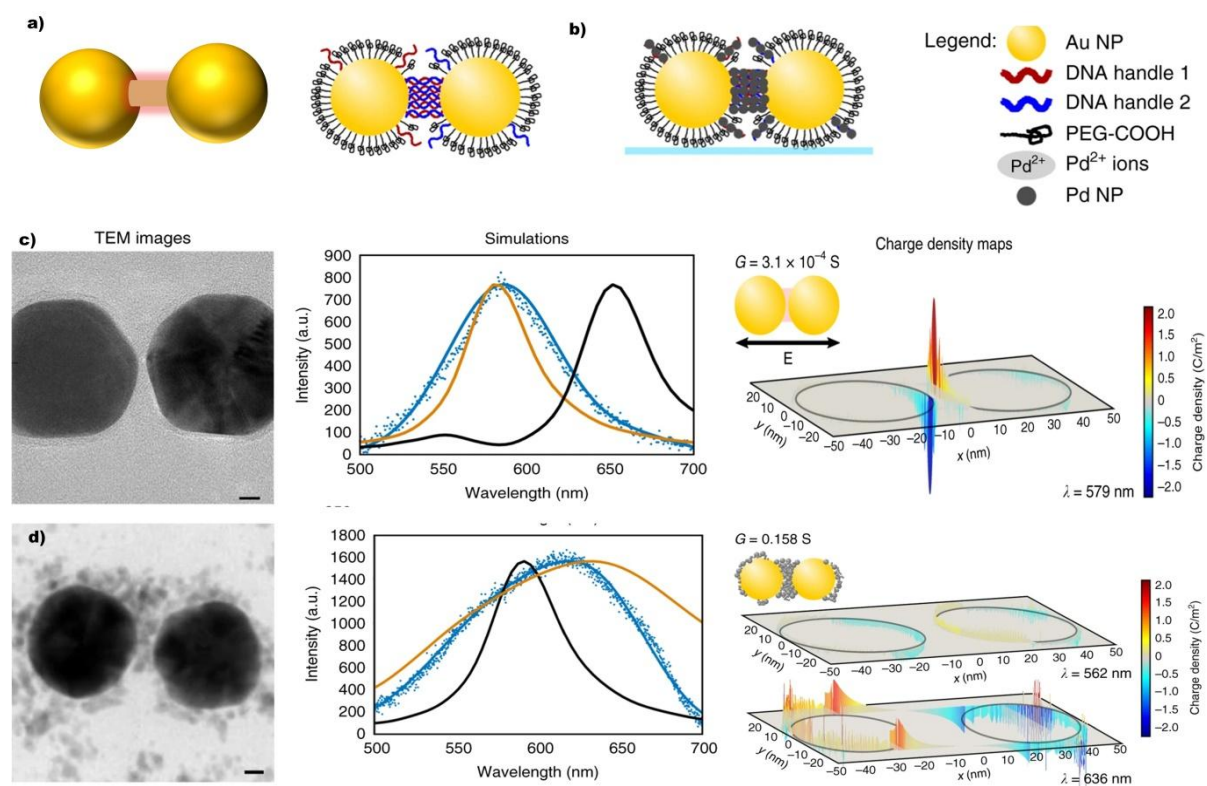

**Figure 3**. Active control of charge transfer in plasmonic nanodimers linked by molecular junctions. Top: Schematic illustrations of (a) quantum corrected model mimicing electrron tunneling through a molecular linker and (b) Au nanoparticle dimer functionalized with DNA molecules incorporated into a monolayer of PEG-COOH (left) and integrated with Pd nanoparticless (right). Bottom: TEM analysis, spectroscopy and charge density maps of Pd nanoparticles in plasmonic nanodimers with gap width of (c) 2 nm and (d) 18 nm. Reproduced under terms of the CC-BY license from ref. [25] Copyright 2018, The Authors, published by Nature Publishing Group.

## 4. Bulk and surface sensing characteristics of charge transfer plasmon resonances

The principle of sensing with charge transfer plasmon resonances was first reported by Perez-Gonzalez *et al*. in their seminal theoretical work.[30] It was found that sharp and ultratunable spectral features of CTP resonances in linked nanodimers can be used for bulk sensing, where they reported high performing CTPR-based sensor with figure-of-merit (FOM) as high as 12.4.[26] Since then, a number of theoretical works have been reported including the most recent work by Dana *et al.* in which they optimized conductively linked asymmetric Au nanodimer for refractive index sensing.[24] They found that, compared to linked symmetric nanodimer, asymmetric nanodimers have higher bulk sensing capability (FOM = 10.88) owing to the excitation of hybridized low-loss modes, like Fano resonances, in such systems. [8, 39]

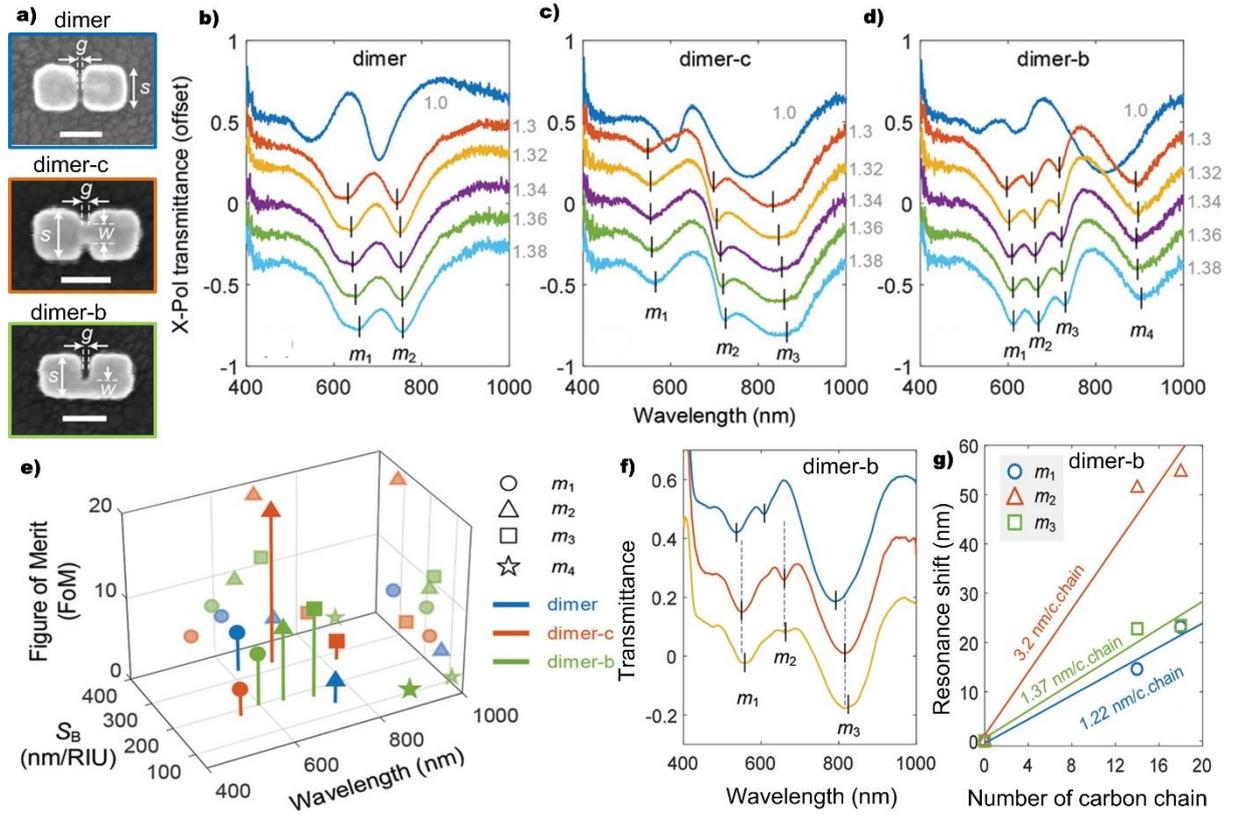

**Figure 4**. Charge transfer plasmon resonance-based bulk and surface sensing. a) Scanning electron micrographs of capacitively coupled sub-5 nm dimer and conductively coupled dimers with center (dimer-c) and bottom (dimer-b) bridges of 40 nm width. Each cubic Au nanodimer has a height of h = 150 nm. Transmission spectra shift as a function of refractive index for b) dimer, c) dimer-c, and d) dimer-b nanostructures. e) Bulk sensing performance of dimer, dimer-c, and dimer-b structures based on cladding index $n_{cl} = 1.3$. f) Transmission spectra and (g) surface sensetivity characteristics of dimer-b as a function of number of carbon chains. Reproduced with permission from ref. [23] Copyright 2021, Wiley-VCH.

Another notable work on CTPR-based sensing has been reported by Tobing *et al.* that demonstrated the emergence of dipole, CTP and hybridized modes in different plasmonic nanostructures including capacitively coupled dimers with sub-5 nm gap and conductively coupled dimers.[23] They designed various plasmonic nanodimers including sub-nm gap unlinked dimers as well as conductively coupled dimers connected with junction bridge positioned at the center (dimer-c) and at the bottom (dimer-b) (see **Figure 4 a**), with the intention of exploiting such nanoarchitectures for bulk and surface sensing applications. As a result of change in the cladding refractive index from n = 1 to n = 1.3, the transmission spectra of these nanostructures can show various modes including longitudinally oscillating dimer-mode ($m_1$) and CTP-like mode ($m_2$) in the dimer; dimer mode ($m_1$), hybrid dimer-CTP mode ($m_2$), and CTP mode ($m_3$) in dimer-c; and dimer mode ($m_1$), hybrid dimer-CTP mode ($m_2$), additional mode ($m_3$), and CTP mode ($m_4$) in dimer-b (See **Figure 4b –d**). In particular, it was found that the conditional emergence of a hybrid dimer-CTP mode of conductively linked dimers based on the cladding refractive index and analyte thickness can present an opportunity for highly specific surface sensing applications (see **Figure 4e-g)**. The aforementioned and other works on charge transfer plasmon resonances imply that these versatile and actively tunable surface plasmon resonances can be exploited for detecting various samples ranging from chemicals and hazardous gas molecules to bulk materials and single molecules (see **Table 1**).

**Table 1.** Charge transfer plasmon resonance-based sensing with different configurations of conductively linked plasmonic nanodimers.

| Dimer Configuration | Gap size | Resonance wavelength | Gap material | Sensitivity [nm/RIU] | FOM | Detection target | Method | Ref. |
| --- | --- | --- | --- | --- | --- | --- | --- | --- |
| Cube dimer | 5 nm | 628 nm | air | 329 [a] | 4.8 | RI | Experimental | [23] |
| Ring-disk dimer | 20 nm | 1760 nm | Au | 840 | 10.88 | RI | Theoretical | [24] |
| Cube dimer | 15 nm | 830 nm | Au | 326 [b] | 18.4 [b] | RI | Experimental | [23] |
| Cube dimer | 15 nm | - | Au | 3.2 [c] | - | Carbon chains | Experimental | [23] |
| Sphere dimer | 5 nm | 839 nm | Pd | - | - | $O_2$ | Theoretical | [38] |
| Sphere dimer | 1 nm | 800 nm | Molecule | 3154.8 | 12.8 | RI | Theoretical | [26] |
| Particle-on-mirror | 1 nm | 665 nm | Molecule | - | - | SAMs | Experimental | [51] |

[a] From two modes that can arise in cube dimer, only the dimer mode ($m_1$) is considered in this case; [b] These data represent sensitivity of dimer-CTP hybrid mode, which arises from the interaction between dimer mode ($m_1$) and CTP mode ($m_3$) as shown in Figure 4; [c] This result denote CTP mode surface sensitivity defined as $S = \Delta\lambda/\Delta t_a$ where, $t_a$ is analyte thickness.

**5. Tunneling charge transfer plasmons in molecular junctions**

Apart from the classical bulk sensing applications, quantum nature of charge and energy flow through conducting linkers is of interest for various applications including understanding the dynamics of plasmon-induced charge transfer, [40-42] photocatalytic water splitting, [43] switchable optical materials, [44] nanoelectronic devices[45] and quantum technology. [46] In particular, the quantum plasmon resonance that appears in molecular tunnel junctions made of two plasmonic nanoresonators bridged by self-assembled monolayers (SAMs) has attracted a growing research interest owing to the opportunities it offers for nano-electronics and single-molecule sensing.[47, 48] In this regard, Tan *et al*. [28] reported direct observation of quantum tunneling between silver nanocube resonators bridged by a SAM made of aliphatic EDT (1,2-ethanedithiolates) and aromatic BDT (1,4-benzenedithiolates) (see **Figure 5a-c)**. The EELS spectra recorded from junctions with SAMs of EDT and BDT reveal that three main plasmon peaks (designated as bonding dipolar plasmon mode (II), transverse corner mode (III), and transverse edge mode (IV)) can be observed along with SAM thickness-dependent, low-energy plasmon mode (I) (**Figure 5d**). Based on the simulation result that shows the transfer of net charge between the Ag cuboiods (see **Figure 5e**), this new mode is assigned as tunneling charge transfer plasmon (tCTP) mode.

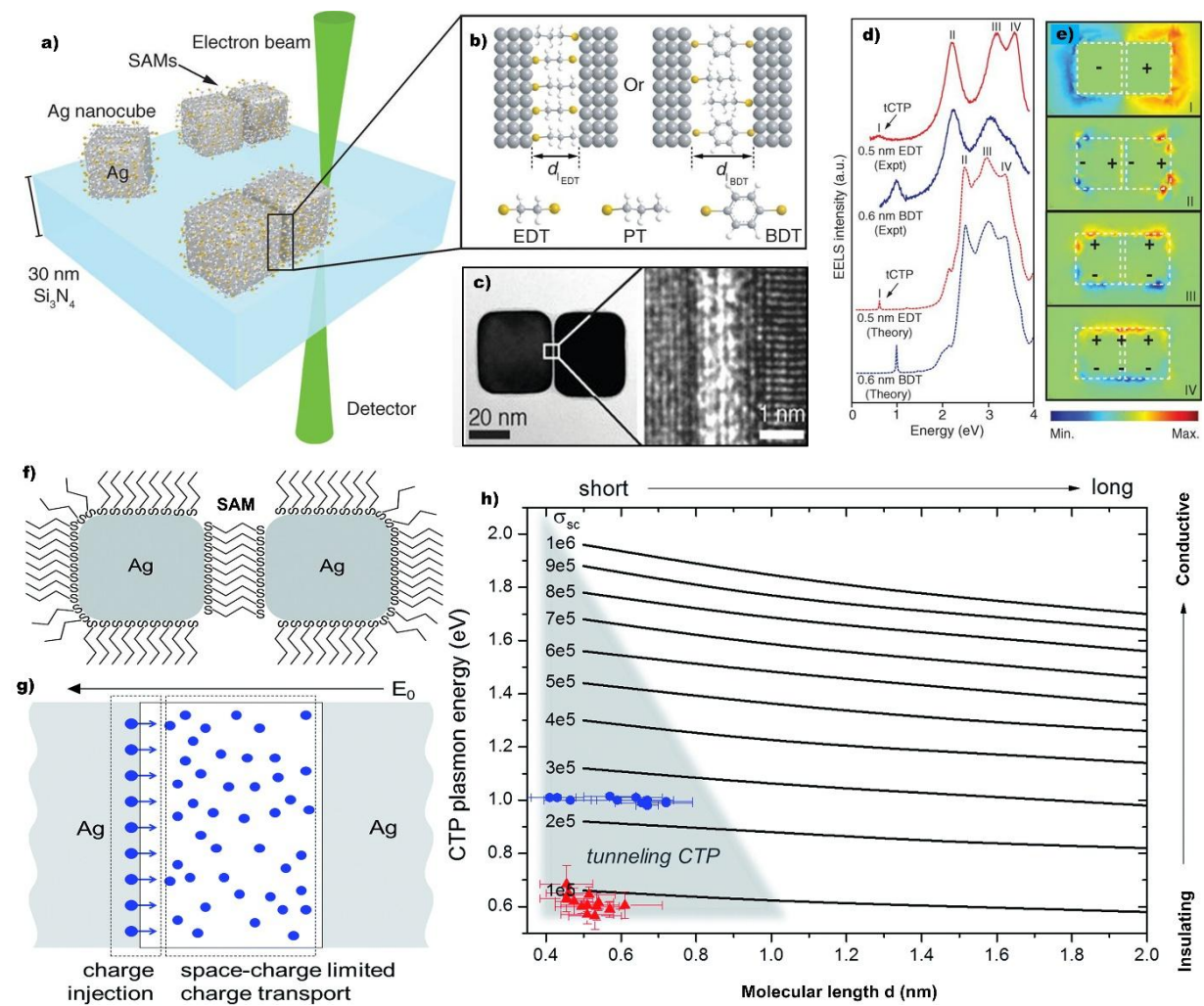

**Figure 5**. Molecular tunnel junction controlled quantum plasmon resonances in plasmonic nanodimers linked by self-assembled monolayers (SAMs). a) Schematic illustration of the molecular tunnel junctions made of two Ag nanoparticles bridged by SAMs of EDT (1,2-ethanedithiolates) and BDT (1,4-benzenedithiolates). (b) Active control of the distance between two adjacent Ag nanoparticles through the EDT and BDT thickness. (c) A high-resolution TEM image of the Ag-SAM-Ag junction. (d) Measured EELS spectra and quantum-corrected simulations of extinction spectra of Ag-SAM-Ag nanosystem, confirming quantum tunneling directly observed as tunneling charge transfer plasmon (tCTP) peak. (e) Simulated electric near-field distributions for the designated plasmonic modes I to IV. Reproduced with permission from ref.[28] Copyright 2014, American Association for the Advancement of Science. (f) Schematic diagram of hybrid Ag–SAM–Ag system. (g) Space-charge corrected electromagnetic model to describe the charge transfer plasmon oscillations of the Ag–SAM–Ag

junction, where negative driving field (during half cycles) induces charge transfer from left to right. (h) Constructed parameter map to correlate resonant CTP plasmon energies (eV), SAM conductivity $\sigma_{sc}$ (S/m), and SAM molecular lengths $d$ (nm) in the Ag–SAM–Ag system for fixed Ag nanocube dimesnions. Reproduced with permission from ref.[48] Copyright 2016, Royal Society of Chemistry.

In related study reported by Wu *et al.*,[48] using the space-charge corrected electromagnetic model treats the charge injection and charge transport separately, they numerically demonstrated establishing a one-to-one relationship between the conductivity of the SAM and the resonant energy of the CTP modes in organic-inorganic hybrid systems made of Ag cube dimer bridged by SAMs consisting of a finite number of molecules (see **Figure 5f - h**). They also estimated the molecular conductance at the CTP resonant frequency for two types of SAMs, where the estimated THz conductance was 0.2$G_0$ per EDT molecule at 140 THz and 0.4$G_0$ for a BDT molecule at 245 THz (where $G_0 = 2e/h$ is the conductance quantum), implying plasmonic oscillations can be used for measuring the THz conductance of single molecules at near-infrared frequencies. [48]

## 6. Molecular conductance switching and sensing in plasmonic junctions

These and other studies on control of charge and energy transfer properties of single-molecule junctions (SMJs) have inspired enormous research interest directing toward single-molecule sensing with the ultimate goal of developing single-molecule opto-electronic devices. [49, 50] To this end, Baumberg *et al.* explored molecular junction conductance of plasmonic nanodimer comprised of Au nanoparticle-on-mirror (NPoM) with actively controlled self-assembled monolayer conductive junction as small as 1.1 nm thick (see **Figure 6a**). [51] They experimentally demonstrated 50 nm blue-shift of CTP as the junction conductance changes from insulating BPT to conductive BPDT (biphenyl-4,4-dithiol) (see **Figure 6b**). Using the LCR model, they also derived a simple analytical description for the blueshifted screened plasmon resonance wavelength shift as [51, 52] $\lambda_{screened} = \frac{\lambda_0}{1+4\varepsilon_d \omega_L^2/\omega_p^2}$, where $\omega_L (= 1/\sqrt{L_g C_s})$ is inductive coupling, $\omega_p (= 2\pi c/\lambda_p)$ is plasma frequency and $\lambda_0$ is unscreened plasmon mode of the NPoM. They demonstrated the conductance of 0.17$G_0$ per BPDT molecule and a total conductance across the junction of about 30$G_0$. By using the number of molecules determined from the conductance-induced blue-shifts, they also obtained surface-enhanced Raman spectroscopy (SERS) of both BPT and BPDT with intensities normalized by the number of molecules (**Figure 6c**). Similarly, Zhang *et al.* applied a combination of mechanically controllable break junction (MCBJ) and in situ SERS methods to investigate single-molecule conductance of prototypical benzene-1,4-dithiol junctions (**Figure 6d&e**). [53]

Plasmon-enhanced break-junction (PEBJ) has also been widely employed for trapping and manipulation of single-molecules with sizes ranging from 2 nm to 1 nm. [54, 55] In particular, Zeng et al. demonstrated direct trapping and in situ sensing of single molecules with sizes down to ~5 Å in solution by employing an adjustable plasmonic optical nanogap and single-molecule conductance measurement.[56]

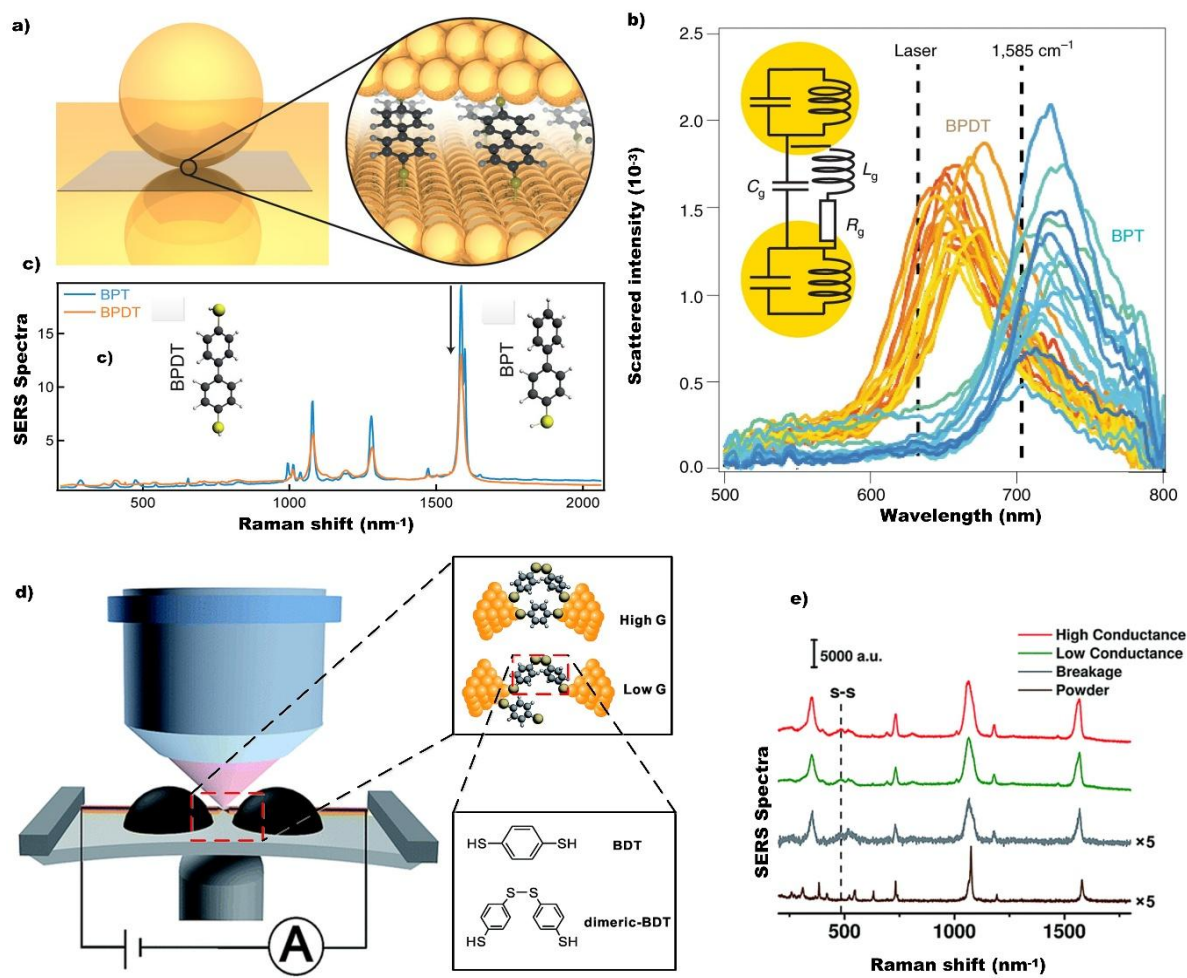

**Figure 6**. Implication of charge transfer plasmons for single-molecule conductance switching and sensing. (a) is placed on a gold film, separated by a thin molecular spacer layer. (b) Molecular layers show conduction-induced blue-shifts in the NPoM scattering resonance for different molecules (here BPDT has lower resistance than BPT). Inset shows the LCR model that accounts for the discharge of gap capacitance $C_g$ when the gap capacitance is shunted by gap resistance $R_g$ and kinetic inductance $L_g$. (c) Average SERS spectra normalized by the number of molecules, comparing SERS intensities from BPDT and BPT. Reproduced with permission from ref. [51] Copyright 2015, American Chemical Society. (d) Schematic illustration of mechanically controllable break junction (MCBJ) SERS setup. f) The hypothetical evolution of the microscopic configuration as the conductance evolves from high conductance to low conductance due to the molecular structures of BDT and dimeric-BDT. g) SERS spectra of BDT collected from different sites of the electrode pair. Inset: the locations of the laser spot. h) SERS spectra collected when the molecular junction was mechanically controlled at the regimes of high conductance (red), low conductance (green), and breakage (grey), respectively. An ordinary Raman spectrum of BDT powder (brown) is displayed for comparison. Laser excitation: 785 nm. Reproduced with permission from ref. [53] Copyright 2018, Royal Society of Chemistry.

## 7. Conclusion

When plasmonic nanoparticles are linked by conductive junction, charge flow can take place across the junction, leading to observation of charge transfer plasmon mode when a minimum conductance threshold is met. These modes are found to be extremely tunable and they have relatively narrow spectral signature, implying their potentials for sensing applications. As a result, CTP has been focus of intense theoretical and experimental research interest for the last ten years. [11, 13, 18, 25, 29] In principle, one can broadly tune the CTP resonances of linked plasmonic nanoparticles by controlling the nanojunction conductance as well as nanoparticle size and shape. And the underlying physics of charge flow across conductive junction are described in terms of classical electrodynamics and quantum mechanical principles. And the spectral shifts of charge transfer plasmons in linked nanoparticles with nanometer scale inter-particle separations can be safely described using the classical universal scaling model. [33] However, when the inter-particle distance enters to the sub-nanometer regime where the non-local screening and electron tunneling effects are prominent, quantum-corrected model is employed. [36] The charge transfer in molecularly linked plasmonic nanodimers can be actively modulated by controlling the junction conductance using self-assembled monolayers, DNA molecules, and Pd nanoparticles, implying their potentials for molecular sensing and switchable opto-electronics.

In this regard, we have explored the sensing capability of CTP and alike modes that arise in strongly coupled and conductively linked plasmonic nanosystems of various configurations. In particular, we have discussed the physical principles and active tuning mechanisms of charge transfer plasmons with possible applications in bulk, surface, and molecular sensing. Beyond biosensing applications, these ultratunable plasmonic modes can also be employed for sensitive and robust detection of gases and chemical reactions. [38, 57, 58] We have also showcased recent theoretical and experimental developments in quantum plasmon resonance that appears in molecular tunnel junctions made of two plasmonic nanoresonators bridged by self-assembled monolayers and their implications for nanoelectronics [59, 60] and single-molecule sensing. We anticipate that these broadly tunable CTP modes can be exploited for realization of ultrasensitive multifunctional plasmonic sensing technologies.

**Acknowledgements**

This work was funded by National Natural Science Foundation of China (Grant No. 62134009, 62121005); and Chinese Academy of Sciences President's International Fellowship Initiative (Grant No. 2023VMC0020).